\documentstyle[twocolumn,aps,epsfig]{revtex}

\title{A device for feasible
fidelity, purity, Hilbert-Schmidt distance and entanglement witness
measurements\\}
\author{R. Filip \\
Department of Optics, Palack\' y University,\\
17. listopadu 50,  772~07 Olomouc, \\ Czech Republic}

\begin{document}
\maketitle
\begin{abstract}
A generic model of measurement device which is able to directly measure
commonly used quantum-state characteristics such as fidelity, overlap,
purity and Hilbert-Schmidt distance
for two general uncorrelated mixed states is proposed. In addition, for
two correlated mixed states, the measurement realizes
an entanglement witness for Werner's separability criterion.
To determine these observables, the estimation of only one parameter --
the visibility of interference, is needed.
The implementations in cavity QED, trapped ion and
electromagnetically induced transparency experiments are discussed.
\end{abstract}

\section{Generic measurement}

A general state in quantum theory is described by density matrix, which
contains all information about the system.
For simple systems, measuring system state $\hat{\rho}$
is well mastered problem, but
with increasing dimension of Hilbert space, it is still
complicated problem.
However, some properties of quantum states can be described in a simpler way
using the parameters, for example: fidelity
$\langle\Psi|\hat{\rho}|\Psi\rangle$ with pure state $|\Psi\rangle$, overlap
$\mbox{Tr}\hat{\rho}^{A}\hat{\rho}^{B}$ between two density matrices, purity
$\mbox{Tr}\hat{\rho}^{2}$ of the state or Hilbert-Schmidt distance
between two different density matrices
$d^{2}(\hat{\rho}^{A},\hat{\rho}^{B})=1/2\times\mbox{Tr}(\hat{\rho}^{A}
-\hat{\rho}^{B})^{2}$.
In this report, we show that these parameters can be measured directly
without state reconstruction even for continuous variable systems.
In recent years, the preparation of quantum states in
QED experiments in high-Q cavities \cite{Brune92}
and trapped-ion experiments \cite{Meekhof96,Liebfried96} are well
mastered. In the last year, an enhancing of nonlinear interaction
between two optical modes in electromagnetically induced transparency
(EIT) \cite{EIT,onephEIT} offers new
possibilities to manipulate with the quantum states.
To measure a value of Wigner function of
corresponding quantum state in a point of phase
space, the effective setup was proposed for all the cavity
QED, trapped-ion and traveling optical pulse experiments \cite{Wigner}.
Therefore, after presenting generic structure of the
measurement we discuss feasible implementations for state of cavity
electromagnetic field, motional degree of freedom of trapped ion and
traveling optical pulses.

We shall employ two density matrices $\hat{\rho}^{A}$ and
$\hat{\rho}^{B}$
for separate systems $0$ and $1$ (generally infinite
dimensional), which can be expressed in particular diagonal
bases $|\psi_{n}\rangle$ and $|\phi_{n}\rangle$
\begin{equation}\label{mixed}
\hat{\rho}^{A}_{0}=\sum_{n}p_{n}|\psi_{n}\rangle\langle
\psi_{n}|,\hspace{0.2cm}\hat{\rho}^{B}_{1}=\sum_{m}r_{m}
|\phi_{m}\rangle\langle
\phi_{m}|.
\end{equation}
The scalar product
$c_{nm}=\langle\psi_{n}|\phi_{m}\rangle$ characterizes the overlap
between two basis states.

We will utilize an auxiliary qubit system $A$
with basis states $|\uparrow\rangle$ and $|\downarrow\rangle$ and
consider two interferometric setups, one for main systems $0$
and $1$ and second for auxiliary qubit system.
For auxiliary system, we need the
unitary transformations $\hat{U}_{H}$
(Hadamard gate)
\begin{eqnarray}\label{transf}
|\uparrow\rangle &\rightarrow& \frac{1}{\sqrt{2}}(|\uparrow\rangle+
|\downarrow\rangle),\nonumber\\
|\downarrow\rangle &\rightarrow& \frac{1}{\sqrt{2}}(|\downarrow\rangle-
|\uparrow\rangle)
\end{eqnarray}
and a phase shift transformation $\hat{U}_{PS}$
\begin{equation}
|\uparrow\rangle\rightarrow\exp(i\psi)|\uparrow\rangle,\hspace{0.5cm}
|\downarrow\rangle\rightarrow |\downarrow\rangle,
\end{equation}
whereas for main systems $0$ and $1$ we need the commonly used
linear coupling operation
\begin{equation}\label{coupl}
\hat{U}_{R}=\exp\left[\frac{\pi}{4}(\hat{a}_{0}^{\dag}\hat{a}_{1}-
\hat{a}_{1}^{\dag}\hat{a}_{0})\right]
\end{equation}
performing the rotation in the phase space,
which is generated by interaction Hamiltonian
\begin{equation}\label{mixi}
\hat{H}_{I1}=i\hbar\xi(\hat{a}^{\dag}_{0}\hat{a}_{1}-\mbox{H.c}),
\end{equation}
where $\xi$ is a linear coupling constant and $\hat{a}_{0}$, $\hat{a}_{1}$
are the annihilation operators of corresponding systems.
The time of interaction should be set to
the value $\tau=\pi/4\xi$. The operations $\hat{U}_{H}$, $\hat{U}_{R}$
and $\hat{U}_{PS}$ can be performed by linear systems and
their implementations obviously do not represent principal problems.

The key point in construction of the proposed measurement is the possibility to
experimentally perform controlled phase shift (CPS) operation
(or an equivalent operation)
\begin{equation}
\hat{U}_{CPS}=\exp\left(i\pi\hat{a}_{0}^{\dag}
\hat{a}_{0}|\uparrow\rangle\langle\uparrow|\right),
\end{equation}
between main system $1$ and auxiliary
qubit system. It can be effectively described by the
following interaction Hamiltonian
\begin{equation}\label{inter}
\hat{H}_{I2}=\hbar\kappa\hat{a}_{0}^{\dag}\hat{a}_{0}|\uparrow\rangle\langle\uparrow|,
\end{equation}
where $\kappa$ is a real coupling constant and the
effective interaction time
is set to be equal exactly to $\pi/\kappa$.
If a state of auxiliary system is
$|\uparrow\rangle$, the above
described operation realizes phase shift about $\pi$ in the ``main''
mode $1$, whereas for state $|\downarrow\rangle$ no phase shift is
induced.

If we are able to perform all the above mentioned operations experimentally,
the proposed measurement can be realized in a way depicted in Fig.~1.
We can see that the device measuring consists of
a main interferometer with the
operations $\hat{U}_{R}$, $\hat{U}_{R}^{-1}$ on systems 0,1 coupled by
CPS-gate to an auxiliary interferometer with the
operations $\hat{U}_{H}$, $\hat{U}_{PS}$ on the qubit system $A$.
We initially assume the qubit system in the state
$|\uparrow\rangle_{A}$.
First, we perform the operations
$\hat{U}_{H}$ and $\hat{U}_{PS}$ on the auxiliary qubit system
and obtain the following state
\begin{eqnarray}
\hat{\rho}&=&\sum_{n,m}p_{n}r_{m}|\Psi_{n,m}\rangle\langle\Psi_{n,m}|,\nonumber\\
|\Psi_{n,m}\rangle&=&\frac{1}{\sqrt{2}}(\exp(i\psi)|\uparrow\rangle+
|\downarrow\rangle)
|\psi_{n}\rangle_{1}|\phi_{m}\rangle_{2}.
\end{eqnarray}
In the second step, we apply the following sequence of transformations
$\hat{U}_{X}=\hat{U}^{\dag}_{R}\hat{U}_{CPS}\hat{U}_{R}$
on the main systems $0$ and
$1$, which, in dependence on the state of auxiliary system,
effectively flips the modes $0$ and $1$ on the output of main systems:
\begin{eqnarray}
\hat{U}_{R}|\psi_{n}\rangle_{0}|\phi_{m}\rangle_{1}|\uparrow\rangle&=&
|\phi_{m}\rangle_{0}|\psi_{n}\rangle_{1}|\uparrow\rangle,\nonumber\\
\hat{U}_{R}|\psi_{n}\rangle_{0}|\phi_{m}\rangle_{1}|\downarrow\rangle&=&
|\psi_{n}\rangle_{0}|\phi_{m}\rangle_{1}|\downarrow\rangle,
\end{eqnarray}
without change of the state of auxiliary system.
If the auxiliary system is in the state $|\downarrow\rangle$, then
the states of modes $1$ and $2$ remain unchanged. On the other hand,
if the qubit is in the state $|\uparrow\rangle$,
then states of systems $1$ and $2$ are flipped.
Thus the state of
the whole system after the interaction is given by the formula
\begin{eqnarray}
\hat{\rho}&=&\sum_{n,m}p_{n}r_{m}|\Psi_{n,m}\rangle\langle\Psi_{n,m}|,\nonumber\\
|\Psi_{n,m}\rangle&=&\frac{1}{\sqrt{2}}(|\downarrow\rangle\,
|\psi_{n}\rangle_{0}|\phi_{m}\rangle_{1}+\nonumber\\
& &+\exp(i\phi)|\uparrow\rangle\,
|\phi_{m}\rangle_{0}|\psi_{n}\rangle_{1}).
\end{eqnarray}
Due to CPS operation, specific entanglement occurs between auxiliary
and main systems.
Finally, the unitary transformation $\hat{U}_{H}$ is performed and
followed by detection $D$ of the states $|\uparrow\rangle$ and
$|\downarrow\rangle$ on auxiliary qubit.

The probabilities $p_{\downarrow}$ and $p_{\uparrow}$
of successful detection of particular state
exhibit oscillatory dependence
on phase $\psi$ and if we assume $p_{\uparrow}$ and $p_{\downarrow}$ for
various $\psi$, then we can estimate the visibility of interference
fringes. After some calculations one can find
that visibility at the output of
an auxiliary interferometer is given by
\begin{equation}\label{vis}
V=\frac{p_{max}-p_{min}}{p_{max}+p_{min}}=
\sum_{n,m}p_{n}r_{m}|c_{nm}|^{2}=\mbox{Tr}\hat{\rho}^{A}\hat{\rho}^{B}.
\end{equation}
Thus, if we consider the independent states (\ref{mixed}) at the input,
then the visibility of interference fringes at the
output of qubit interferometer is given exactly
by the overlap $O=\mbox{Tr}\hat{\rho}^{A}\hat{\rho}^{B}$.
We can define overlap observable $\hat{O}$
on the total system in the following form
\begin{eqnarray}\label{observ}
\hat{O}&=&\sum_{n',m'}|\psi_{n'}\rangle_{00}\langle\psi_{m'}|\otimes
|\psi_{m'}\rangle_{11}\langle\psi_{n'}|,\nonumber\\
O&=&\mbox{Tr}\hat{O}\hat{\rho},
\end{eqnarray}
where $\hat{\rho}=\hat{\rho}^{A}\hat{\rho}^{B}$ is the input density matrix.
Note, that operator $\hat{O}$  is exactly
flip operator \cite{Werner89,Horodecki96}
making transformation $\hat{O}(\psi\otimes\phi)=\phi\otimes\psi$.
Particularly, the scalar product
$|_{0}\langle\Psi|\Phi\rangle_{1}|^{2}$ can be
measured for two pure states and fidelity
$F=_{1}\langle\Psi|\hat{\rho}_{0}|\Psi\rangle_{1}$ for one pure and
one mixed state. It is important that only single parameter (the
visibility) has to be estimated to measure overlap,
irrespective of the complexity of the state of main systems.

To this moment, we have assumed that the two systems $0$ and $1$ are initially
uncorrelated and thus the total state was considered in separate form
$\hat{\rho}=\hat{\rho}^{A}_{0}\hat{\rho}^{B}_{2}$.
For general state written in any local bases
$|\psi_{n}\rangle_{0}$ and $|\phi_{m}\rangle_{0}$
\begin{equation}
\hat{\rho}=\sum_{n,m,k,l}\rho_{nmkl}|\psi_{n}\rangle_{00}\langle\psi_{k}|\otimes
|\phi_{m}\rangle_{11}\langle\phi_{l}|
\end{equation}
it can be found after some algebra that the measured visibility is
\begin{equation}\label{entvis}
V=\left|\sum_{n,m,k,l}\rho_{nmkl}\langle\phi_{l}|\psi_{n}\rangle\langle\psi_{k}|
\phi_{m}\rangle\right|,
\end{equation}
which can be re-arrangered in a way
\begin{equation}
V=\left|\langle\Lambda|\hat{\rho}^{T_{2}}|\Lambda\rangle\right|.
\end{equation}
Here, the partial transposition is defined as follows
\begin{equation}
\rho^{T_{2}}_{mn,m'n'}=\langle m|_{0}\langle n|_{1}\hat{\rho}^{T_{2}}
|n'\rangle_{1}|m'\rangle_{0}=\rho_{mn',m'n}
\end{equation}
and $|\Lambda\rangle$ is an unnormalized maximally entangled state
\begin{equation}
|\Lambda\rangle=\sum_{j}|\phi_{j}\rangle_{0}|\phi_{j}\rangle_{1}.
\end{equation}
The measurement suggested above can be considered as
measuring the observable $\hat{O}$ in (\ref{observ}) or
reformulating it as positive operator values measure (POVM)
in the following form
\begin{eqnarray}\label{POVM}
V&=&|\mbox{Tr}\hat{\rho}\hat{V}|,\hspace{0.3cm}
\hat{V}=\hat{\Pi}_{+}-\hat{\Pi}_{-},\nonumber\\
\hat{\Pi}_{+}&=&\sum_{n\geq m} |+,m,n\rangle\langle +,m,n|,\nonumber\\
\hat{\Pi}_{-}&=&\sum_{n> m} |-,m,n\rangle\langle -,m,n|,\nonumber\\
|+,m,n\rangle&=&\frac{1}{\sqrt{2}}(|\psi_{n}\rangle_{0}|\psi_{m}\rangle_{1}+
|\psi_{m}\rangle_{0}|\psi_{n}\rangle_{1}),\nonumber\\
|-,m,n\rangle&=&\frac{1}{\sqrt{2}}(|\psi_{n}\rangle_{0}|\psi_{m}\rangle_{1}-
|\psi_{m}\rangle_{0}|\psi_{n}\rangle_{1}).
\end{eqnarray}
The particular POVMs $\hat{\Pi}_{+}$ and $\hat{\Pi}_{-}$
represent projectors onto symmetric (antisymmetric) subspaces of the total
space of systems 0 and 1. Hence $\hat{V}$ is dichotomic variable
with eigenvalues $\pm 1$.

This overlap measurement can be used to quantify another
important state characteristics.
Direct application is measuring of probability $p
=\langle\Psi|\hat{\rho}|\Psi\rangle$, if the pure state $|\Psi\rangle$
can be generated. It can be important in the cases, when a measuring
apparatus consisting of all the projections cannot be realized,
however the particular eigenstates can be prepared.
Another possible applications are measuring of purity
$P=\mbox{Tr}\hat{\rho}^{2}$ and linear entropy
$S_{L}=1-\mbox{Tr}\hat{\rho}^{2}$.

In addition,
the proposed measurement is nondemolition overlap measurement.
It can be simply proved:
after measurement the output state is described as balanced mix
of input density matrices
\begin{equation}\label{output}
\hat{\rho}_{\rm out}=\frac{\hat{\rho}_{1}^{A}\hat{\rho}_{2}^{B}+
\hat{\rho}_{1}^{B}\hat{\rho}_{2}^{A}}{2}
\end{equation}
and if one applicates once more the same proposed
measurement, then one obtain the same visibility (\ref{vis}).
It can be verified by inserting Eq.~(\ref{output}) to
Eq.~(\ref{entvis}). The nondemolition character can be utilized
to measure commonly used Hilbert-Schmidt distance
\begin{equation}
d^{2}(\hat{\rho}^{A},\hat{\rho}^{B})=\frac{P^{A}+P^{B}}{2}-O^{AB}
\end{equation}
between two states of the separated systems.
To measure $d^{2}$, we can first
perform the measurement of particular purities $P^{A}$ and
$P^{B}$ and then use the same ensemble to measure
the overlap $O^{AB}$ between systems $A$ and $B$.

It is interesting that the visibility is given only by positive (but not
completely positive) transformation on the density matrix, which
can be connected with some entanglement witness \cite{Terhal00} using
Werner-Peres-Horodecki's (WPH) criterion
\cite{Werner89,Peres96,Horodecki96}. It can be interesting to analyse
how the visibility depends on the
correlations between cavities. We shall focus on the difference of
probabilities $\Delta=p_{\uparrow}-p_{\downarrow}$ instead of the
visibility $V$. First, we use the setup without $CPS$ gate and
fix the phase $\psi$ in such a way that $p_{\uparrow}=1$.
Then after performing complete measurement, the difference of
probabilities
is given by $\Delta=\langle\Lambda|\hat{\rho}^{T_{2}}|\Lambda\rangle$ and can be
both negative or positive. Due to WPH criterion of
separability, if $\Delta$ is negative
one can be sure that the total state is entangled.
Thus the above proposed measurement can be considered as the entanglement
witness, considering POVM (\ref{POVM}) as the definition of measurement.
The discussion about a set of states showing the inseparability by this
measurement is still in the investigation.
Instead, we present here the simplest illustrative
example with qubit systems:
one can simply find that for entangled
state $1/\sqrt{2}\times (|01\rangle-|10\rangle)$ in systems 0 and 1,
the visibility is $V=1$, whereas
for maximally classically correlated state $1/2\times (|01\rangle\langle
01|+|10\rangle\langle 10|)$ the visibility is vanishing, similarly as for
orthogonal states $|0\rangle$, $|1\rangle$. Thus, an appropriate entanglement in the total system
can lead to the increase of the visibility in measuring device.

\section{Possible experimental realizations}

The simplicity of the proposed measurement, based on interferometric
experiments, gives a possibility to experimentally implement the scheme in
atom-field cavity and trapped ions experiments. In both these areas of
quantum optics, the interferometric experiments are well mastered and
the desired interaction (\ref{inter}) can be achieved with sufficiently large
efficiency. Recently due to new experimental results in
EIT enhancement of Kerr
nonlinear interaction needed to construct CPS-gate, it seems to be
reasonable to consider an implementation in this field too.
Now, we shortly point out the basic principles
of experimental implementations.

In atom-field cavity experiments, to measure the overlap between light
fields in different cavities, an experimental setup previously used
by Raimond and co-workers \cite{Raimond97} can be utilized.
The scheme is sketched in Fig.~1. A stream of two-level
atoms serves as an auxiliary system
and electromagnetic fields in cavities $C_{0}$ and
$C_{1}$ are the systems of interest.
The cavities are coupled through a
controlled superconducting waveguide to perform a linear interaction (\ref{coupl}).
Thus the main interferometer is realized by the coupling
between cavities, whereas Ramsey atomic interferometer is used for
auxiliary system.
We will neglect relaxation processes in the cavities, as
well as for the atoms, during the experiment. This is
realistic for experimentally achievable cavity quality factors of a few
$10^{9}$ corresponding to photon lifetimes of the order a few ms
\cite{Raimond97}. The principle of phase manipulation with field in
cavity was previously described in \cite{Brune92,Davidovich96} and shows
that three-level atoms interacting with light in cavity (in large detuning
limit) can be effectively described by the Hamiltonian (\ref{inter}) with
$\kappa=\Omega/\delta$, where $\Omega$ is Rabi vacuum coupling
and $\delta$ is detuning between atomic and cavity frequency.
As in the experiment \cite{Raimond97},
the coupling between $C_{0}$ and $C_{1}$ plays no role
during the interaction (\ref{inter}),
provided that the interaction time is much
shorter than coupling period and it is much smaller than
photon lifetime in cavity.

The measurement procedure can be performed as follows. First, the
coupling between $C_{0}$ and $C_{1}$ is switched on. After $50:50$
energy exchange between the cavities $C_{0}$ and $C_{1}$, the atom effusing
from an oven O, velocity selected in zone V, is excited into a circular
Rydberg state in zone B. The atom is prepared, before entering $C_{0}$, in
superposition $\frac{1}{\sqrt{2}}(\exp(i\psi)
|\uparrow\rangle+|\downarrow\rangle)$
by a classical microwave field applied in zone $R_{1}$. Then the
interaction (\ref{inter}) in cavity realizes CPS
operation. Another $50:50$
cavity coupling follows and the atomic states are rotated
in a second classical microwave zone $R_{2}$ (performing again the
transformation (\ref{transf})). We choose here different phases for
the classical fields in $R_{1}$ and $R_{2}$ to prepare superposition
with variable phase $\psi$.
The atom is finally detected by field-ionization
counters $D_{e}$ end $D_{g}$, either in state
$|\uparrow\rangle$ or in state $|\downarrow\rangle$.
The measurement accuracy depends on the detector's selectivity, that is,
the ability to distinguish between the two atomic states and on
the velocity spread of the atomic beam.
Repeating the procedure with different phases,
visibility of interference in the Ramsey interferometric setup can
be determined. As was shown above, this is equal to the overlap between states
in cavities $C_{0}$ and $C_{1}$.


Analogical possibility of experimental demonstration of the proposed
measurement arises in trapped ion experiments.
The setup can be implemented to measure
the overlap between two vibrational degrees
of freedom of trapped ion.
We consider two-dimensional trapped-ion model, where
the ion is confined in a linear rf Paul-trap and
is irradiated by two classical laser beams propagating along
the $x$ (system 0) and $y$ (system 1) directions.
Due to harmonic potential of the trap and
interaction with laser beams, a complicated
trapped ion structure can be approximated by
two-dimensional harmonic oscillator in
$x$,$y$-vibrational degrees of freedom coupled to two-level electronic system, which
will be treated as the auxiliary qubit.

The construction of the CPS gate in trapped ion
implementation is similar to commonly used QND
measurement of vibrational energy \cite{QND}. The QND interaction
is only excited between
$x$-vibrational degree of freedom and internal electronic states.
The $y$-motional mode must not be changed by this interaction.
In the resonant approximation between Raman transition and
Stark-shifted electronic levels and using second-order
Lamb-Dicke limit, a kind of interaction Hamiltonian
\begin{equation}
H_{I}=\hbar\Omega\hat{a}_{x}^{\dag}\hat{a}_{x}
(|\uparrow\rangle\langle\downarrow|+|\downarrow
\rangle\langle\uparrow|)
\end{equation}
can be obtained between
vibrational harmonic oscillation in $x$ direction and
an internal electronic states $|\uparrow\rangle$ and $|\downarrow\rangle$.
The interaction time must be adjusted to $t=\pi/2\Omega$.
Due to different interaction Hamiltonian, we must modify the setup
in equivalent manner. Instead of basis states $|\uparrow\rangle$,
$|\downarrow\rangle$ of auxiliary qubit system, we will use the
interaction invariant basis states
$|+\rangle=1/\sqrt{2}\times(|\uparrow\rangle+|\downarrow\rangle)$, $|-\rangle=
1/\sqrt{2}\times(|\uparrow\rangle-|\downarrow\rangle)$ in auxillary system with
corresponding change of operation (\ref{transf})
and introduce additional phase-shift transformation
$\hat{U}_{PS}=\exp(i\pi/2\times\hat{a}^{\dag}_{x}\hat{a}_{x})$
after CPS-gate in motional-$x$ mode
to retrieve correct balance between phases. Then the measurement can be
performed in a way analogous to that proposed previously.
To realize a linear mixing (\ref{coupl}) of the $x$ and $y$ modes,
we can use two classical fields
with an appropriate mutual detuning, as it was proposed in work
\cite{Parkins00}. Then,
in the first order of Lamb-Dicke approximation and using rotating-wave
approximation, the effective Hamiltonian between $x$ and $y$ directions
of vibrational modes has the form (\ref{mixi}).
The operations with vibrational auxiliary system can be performed
using stimulated Raman pulses, similarly to
ion-interferometry experiments \cite{Poyatos96}.
To detect lower internal state
$|\downarrow\rangle$ we can apply Doppler beam and measure the
fluorescence \cite{Monroe95} (for the experiment reported in
\cite{Monroe96}, the detection efficiency is close to unity).
This implementation illustrates that the generic measurement setup can
easily be modified for various experimental situations.


Recently, the enhancement of nonlinear cross-Kerr interaction between
two pulses using EIT was performed.
It is possible to change the phase of
light pulse about $\pi$
by the single photon pulse \cite{EIT}.
To realize a large
cross phase modulation on a single photon level, both cavity
and free-medium regimes have been considered
\cite{onephEIT}.
To perform CPS-gate we can use this interaction with effective Hamiltonian
$\hat{H}=\hbar\kappa\hat{n}_{0}\hat{n}_{A}$, where $\hat{n}_{0}$ is
number operator of photons in mode 0 and $\hat{N}_{A}$ is number operator
of photons in
auxiliary mode $A$. The auxiliary two-level system is represented by two
paths of a photon in Mach-Zehnder interferometer $A$ and the transformation $R$
could be constructed using the beam splitters $BS$ and phase-shifters
$PS$ in such a way, that the total scheme,
depicted for traveling-wave configuration
in Fig.~3, performs the required
overlap measurement. The systems $0$ and $1$ are represented by
different modes of optical pulses in the
upper interferometer and their states are mutually
exchanged at the output, if the photon is presented in upper arm of
auxiliary
interferometer $A$. The visibility is determined from the probabilities
of photon detection on the detectors $D$. However, some technical problems
can occur in realization of this setup, in addition to preparation of intensive
Kerr-like interaction, with obtaining efficient one photon source and
synchronization of pulses interacting in EIT medium.

\medskip
\noindent {\bf Acknowledgments}
The author would like to thank J. Fiur\' a\v sek for the important
suggestions about the interpretation and reading of the paper
and M. Du\v sek, M. \v Sa\v sura and Prof. J. Pe\v rina for
stimulating discussions.
That result was supported by the project LN00A015 and CEZ:J14/98
of the Ministry of Education of Czech Republic.

\begin{figure}
\vspace{0cm}
\caption{Generic measuring device: H -- Hadamard gates, CPS -- controlled
phase shift, PS -- phase shift, $R$ -- coupling gate, D -- detector.}
\label{schema0}
\end{figure}
\begin{figure}
\vspace{0cm}
\caption{Measuring device (atom-field cavity implementation): O -- oven,
V -- velocity selection, B -- excitation box, $R_{1}$, $R_{2}$ --
Ramsey zones, $C_{0}$, $C_{1}$ -- high-Q cavities, $D_{e}$, $D_{g}$ --
ionization detectors.}
\label{schema1}
\end{figure}
\begin{figure}
\vspace{0cm}
\caption{Measuring device (EIT implementation): BS -- beam splitters, PS
--phase shifters, KERR -- nonlinear medium with EIT, D -- detectors.}
\label{schema2}
\end{figure}


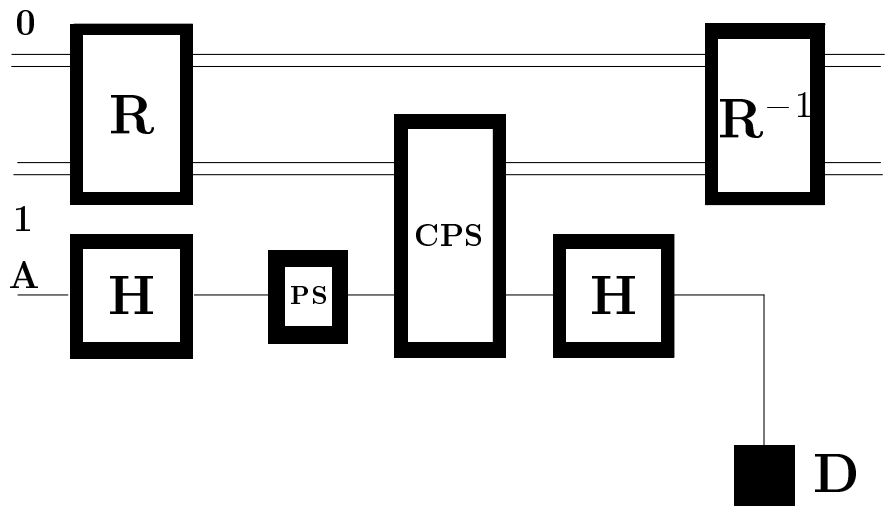
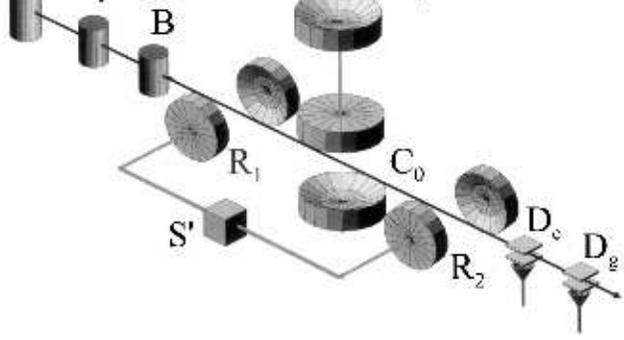
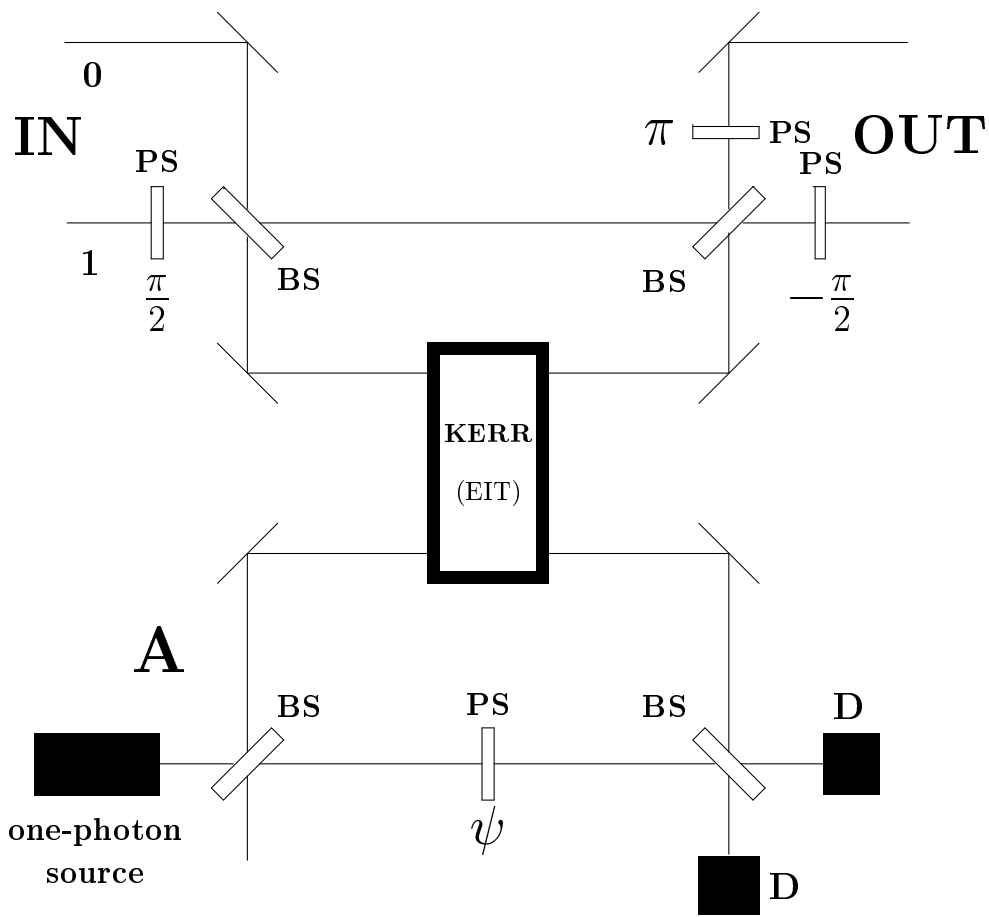
\end{document}